\documentclass[doublecol,amsmath,amssymb,figures]{epl2}
\usepackage{graphicx}

\newcommand{\bc}{\begin{center}}
\newcommand{\ec}{\end{center}}
\newcommand{\be}{\begin{equation}}
\newcommand{\ee}{\end{equation}}
\newcommand{\ba}{\begin{array}}
\newcommand{\ea}{\end{array}}
\newcommand{\beq}{\begin{eqnarray}}
\newcommand{\eeq}{\end{eqnarray}}

\begin{document}

\title{Nonequilibrium critical relaxation at a first-order phase transition point}
\shorttitle{Nonequilibrium critical relaxation at a first-order phase transition point}

\author{
Michel Pleimling\inst{1} \and Ferenc Igl\'oi\inst{2,3}}
\shortauthor{Michel Pleimling and Ferenc Igl\'oi}
\institute{
  \inst{1} Department of Physics, Virginia Polytechnic Institute and State University,
 Blacksburg, Virginia 24061-0435, USA \\
  \inst{2} Research Institute for Solid State Physics and Optics, H-1525 Budapest, Hungary\\
  \inst{3} Institute of Theoretical Physics, Szeged University, H-6720 Szeged, Hungary
}

\pacs{64.60.Ht}{Dynamic critical phenomena}
\pacs{75.40.Gb}{Dynamic properties (dynamic susceptibility, spin waves, spin diffusion, dynamic scaling, etc.)}
\pacs{05.70.Fh}{Phase transitions: general studies}


\abstract{
We study numerically the nonequilibrium dynamical behavior of an Ising model with mixed
two-spin and four-spin interactions after a sudden quench from the
high-temperature phase to the first-order phase transition point. The autocorrelation function
is shown to approach its limiting value, given by
the magnetization in the ordered phase at the transition point, $m_c$, through a stretched
exponential decay. On the other hand relaxation of the
magnetization starting with an uncorrelated initial state with magnetization, $m_i$, approaches
either $m_c$, for $m_i>0.5$, or zero, for $m_i<0.5$. For small $m_i$ and for $m_i$ slightly larger
than $0.5$ the relaxation of the magnetization shows an asymptotic power-law time dependence,
thus from a nonequilibrium point of view the transition is continuous.
}


\maketitle

\section{Introduction}

Most of our knowledge about the properties of phase transitions has accumulated for
continuous or second-order transitions for which the concepts of scaling and universality, as well
as the application of the method of renormalization group, have provided a deep understanding \cite{fisher}.
Considerably less is known about singularities at first-order phase transitions, although in nature
this type of transition is very common \cite{binder}. At a
first-order phase transition point one observes phase co-existence where several thermodynamical quantities,
such as the internal energy and the order parameter (magnetization), have a discontinuity, whereas
the correlation length is generally finite. In spite of this finite correlation length 
some response functions, such as the magnetic susceptibility, have a scaling behavior in a finite
system\cite{fisherberker}
that involves the discontinuity fixed-point exponent $y_d=d$ where $d$ denotes the dimension
of the system \cite{disc}. From a dynamical point of view the relaxation time
is finite at a first-order transition point, yielding equilibrium autocorrelations and relaxation 
functions that decay exponentially.

Many studies have been devoted to the investigation of the formation of the ordered phase
when the temperature is lowered below the phase transition point, $T_c$. Thus, the
nonequilibrium dynamics of the nucleation process\cite{gunton} and the morphology of the
solidification\cite{grana}
are key problems in material science. From a more theoretical perspective one
is interested in the coarsening process \cite{bray} that takes place when the system is suddenly quenched from the high-temperature
phase at or below the phase transition point. If the quench is performed below $T_c$, then
the competition between the stable ordered phases leads to the same type of qualitative picture
independently of the order of the transition at $T_c$ \cite{Lorenz07}. If, however, we
quench the system at the very transition point, important differences are expected in the two cases
due to the different behavior of the correlation length.

In the past this type of phenomena, i.e. nonequilibrium critical relaxation, has been thoroughly studied in the case of a second-order transition point \cite{jans92,cg04}, whose investigations were also motivated by
the appearence of ageing. On the contrary only little attention has been paid to the case when the
transition is of
first order. It has been demonstrated that nonequilibrium relaxation starting from an ordered or a
mixed-phase initial state can be used to locate accurately the phase transition point and to decide
about the order of the transition \cite{zheng1,ozeki}. For models with quenched disorder the
tricritical value of the
dilution has also been studied by this method \cite{zheng2}. However, we are not aware of any study in which the
singularities in the dynamical quantities are systematically investigated. In this paper we perform
an investigation of the nonequilibrium dynamics at a first-order phase transition that addresses
questions well established for a second-order transition point.
In particular we consider the asymptotic behavior of the autocorrelation function as well
as the relaxation of the order parameter after a sudden quench from the high-temperature phase to the
first-order transition point.

The structure of the paper is the following. After a brief recapitulation of the known results of nonequilibrium
critical relaxation we present a detailed numerical study performed on a square lattice Ising model with mixed
two-spin and four-spin product interactions \cite{turban1,turban}.
This system has a first-order phase transition and
the location of the transition point is known by duality. The numerical results obtained in this
model are consistent with the existence of a divergent nonequilibrium relaxation time. We interpret
its origin and argue that the observed nonequilibrium dynamical behavior is generic at first-order
phase transitions.

\section{Nonequilibrium critical dynamics: a reminder}
At a second-order transition point both the correlation length, $\xi$, and the
relaxation time, $\tau$, are divergent, and they are related through: $\tau \sim \xi^z$.
The dynamical exponent, $z$, which generally depends on the local dynamics,
on conservation laws, and on symmetries, is enough to classify the dynamical universality class at
equilibrium \cite{hh77}. In a nonequilibrium situation, when the system is
quenched from the high-temperature phase as an initial state to the critical point, new nonequilibrium
exponents have to be defined \cite{jss89,huse89}. The reason for this is
the broken time translation invariance due to a discontinuity at the
time horizon ("time-surface"). As a result the autocorrelation function, $C(s,t)$, is
generally non-stationary: it depends on both the waiting time, $s$, and the observation time, $t$.
In the limit $t \gg s$ we have asymptotically: $C(s,t) \sim
t^{-\lambda/z}$, which involves the nonequilibrium autocorrelation exponent $\lambda$ \cite{huse89}.
In another related process the initial state is prepared with a
small, non-vanishing magnetization, $m_i$, and one measures its relaxation,
which for short times behaves as $m(t) \simeq m_i t^{\theta}$. Here the
initial slip exponent $\theta$ satisfies the scaling relation:
$\lambda=d-\theta z$ \cite{jss89,jans92}. We note that in mean-field theory $\theta=0$, whereas in
real systems the exponent $\theta$ is generally larger than zero
and can be expressed as $\theta=(x_i-x)/z$ where $x=\beta/\nu$ is
the anomalous dimension of the bulk magnetization ($\beta$ and $\nu$ are the standard magnetization
and correlation length exponents, respectively) and $x_i$ is called the anomalous dimension of
the initial magnetization.

\section{Ising model with multispin interaction}
At a first-order transition point the correlation length stays finite, $\xi < \infty$, and the same is
true for the equilibrium relaxation time, $\tau < \infty$. The magnetization displays a jump
from zero above $T_c$ to some value $m_c > 0$ below $T_c$.
In this case the static critical exponents are formally given by $x=\beta=0$ and $\nu=1/d$, where
the latter is the discontinuity fixed point value \cite{disc}. The specific model we study numerically in the
following belongs to a class of square lattice Ising models with two-spin interactions in the vertical
direction and $n$-spin product interactions in the horizontal direction that
is defined by the Hamiltonian
\be
H=-\sum_{i,j}\left(J_2 \sigma_{i,j} \sigma_{i,j+1} + J_n \prod_{k=0}^{n-1} \sigma_{i+k,j}\right)\;.
\label{hamilton}
\ee
The model is self-dual \cite{gruber,turban1,turban} and the self-dual point is located at: $\sinh(2J_2/k T_c)\sinh(2J_n/k T_c)=1$.
For $n=2$ we of course recover the standard Ising model. Whereas for $n=3$ the transition is found to be
continuous and to belong to the universality class of the four state Potts model
\cite{n3t,n3a,n3b,n3c,n3d,n3e}, for $n \ge 4$
the transition is of first order \cite{n4,blote}.
In the following we consider the model with four-spin interaction, in which case the ordered phase has
eightfold degeneracy\cite{eight} and the phase transition of the system is comparable with that in the eight-state
Potts model.
If one takes $J_2=J_4$, numerical results \cite{blote} indicate a latent heat of $\Delta/k T_c=0.146(3)$ and a
jump of the magnetization from zero to $m_c=0.769(6)$. The correlation length has not been measured in this system,
but from the snapshots of the simulations one expects somewhat different correlation lengths in the
horizontal and in the vertical directions that are both of the order of ten lattice constants.


\begin{figure*}
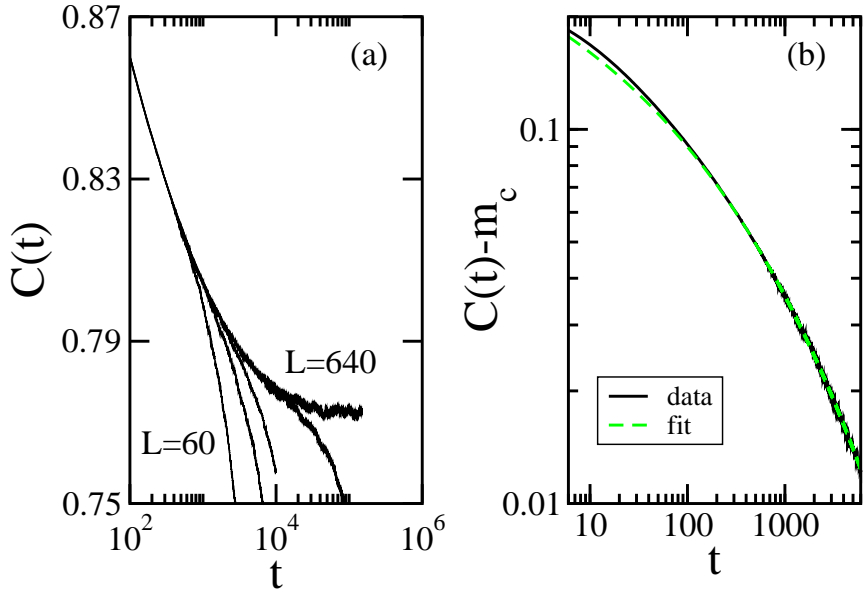

\onefigure[scale=0.6]{relax_new_fig1.eps}
\caption{
\label{auto}
(a) The autocorrelation as a function of time for systems composed of $L^2$ spins
with $L = 60$, 90, 120, 240, and 640. Three different time regimes can be identified, see main text.
(b) The connected autocorrelation function $C(t)-m_c$ as a function of time. A stretched exponential
decay (dashed line) is observed for times $t > \tau$ with $\tau \approx 200$.
 }
\end{figure*}

\section{Single spin autocorrelation function}
The single spin autocorrelation function of the model is defined as $C(s,t)=\langle \sigma_{i,j}(s)\sigma_{i,j}(t)\rangle$, which is then averaged over all spins, $1 \le i,j \le L$,
and periodic boundary conditions are applied in both directions. Starting with a disordered
initial state, which corresponds to $T=\infty$, we apply non-conserving spin-flip dynamics
at the temperature $T=T_c$. In the
following we consider no waiting time, i.e. $s=0$, and use the notation: $C(0,t) \equiv C(t)$.
In order to monitor finite-size effects we simulated systems containing $L^2$ spins with $L$ ranging from
60 to 640. For the largest size we averaged over 2000 different samples with different realizations
of the noise, more samples were simulated for smaller system sizes. As usual, time is measured in
Monte Carlo steps where on average every spin is flipped once per step.
The calculated autocorrelation functions for different finite systems 
are plotted in Fig. \ref{auto}. Rather large systems are needed in the calculation
in order to handle the finite-size effects. From Fig. \ref{auto}a we can identify three different time-regimes.
i) For short times, which are shorter than the equilibrium relaxation time, $t<\tau$, there is a
fast decay of $C(t)$, which is described by a stretched exponential function. ii) This initial decay is followed by a relatively slow approach to the
equilibrium limiting value $\lim_{t \to \infty} \lim_{L \to \infty} C(t)=m_c$. Indeed the plateau
of $C(t)$ in Fig.\ref{auto}a gives an accurate estimate $m_c=0.770(3)$ in accordance with previous
numerical results \cite{blote}. For the largest finite system we
have checked the behavior of the connected autocorrelation function and we have obtained
a stretched exponential form:
$C(t)-m_c \sim \exp(-bt^a)$, with an exponent $a=0.198(3)$, see Fig. \ref{auto}b. We note that in
the short-time regime the decay is described by a somewhat larger effective exponent in the
stretched exponential, $\tilde{a} \approx 0.32$. iii) The third regime
of $C(t)$ is seen in finite systems\cite{diehl0}, in which it starts
to decrease from the saturation value after passing a cross-over time: $t_L \sim L^z$.
The dynamical exponent measured in this part of the figure is found to be $z=2.05(10)$,
which is consistent with $z=2$.

At this point we can conclude that for large systems
the nonequilibrium autocorrelation function at a first-order transition point approaches
the bulk magnetization, $m_c$. Thus although the average magnetization of the system is vanishing
(see the next section) the dynamical correlations display a finite asymptotic value. On the
other hand the connected autocorrelation function has an asymptotic
stretched exponential decay.

\section{Relaxation of the magnetization}
Next we investigate the relaxation of the magnetization at $T_c$ where we start with an uncorrelated
initial state with a magnetization $m_i \le 1$. 
In order to obtain reliable data for this noisy quantity, we averaged over
typically 250000 different samples with different initial states and different realizations of the noise.
The data discussed in the following have been obtained for systems containing $240^2$ spins. 
We checked that the presented data are free of finite-size effects by simulating some larger systems
for selected values of the initial magnetization $m_i$.


\begin{figure*}
{\par\centering \resizebox*{0.5\textwidth}{!}{\includegraphics{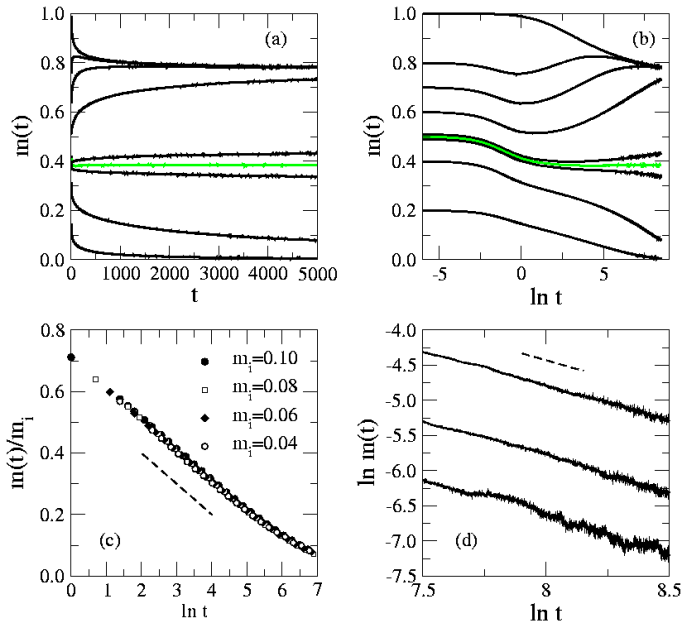}} \par}
\caption{
\label{relax}
(a) Time evolution of the magnetization when starting from a state with an initial
magnetization $m_i$, with $m_i = 1$, 0.8, 0.7, 0.6, 0.51, 0.5, 0.49, 0.4, and 0.2 (from top to bottom).
The grey line separates the two different fixed point. (b) The same as (a), but now the magnetization is
shown as a function of $\ln t$, thus illustrating the non-monotonic behavior that one may encounter at 
early times. (c) $m(t)/m_i$ as a function of $\ln t$ for different small values of $m_i$. The initial decay 
is given by $0.7 - 0.1 \ln t$. The slope of the dashed line is 0.1. (d) At later times the decay
for small initial magnetization is given by a power-law with an exponent $\theta' \approx -1$ (dashed line).
Shown are data obtained for $m_i =0.2$, 0.1, and 0.05 (from top to bottom).
}
\end{figure*}

\begin{figure*}
{\par\centering \resizebox*{0.5\textwidth}{!}{\includegraphics{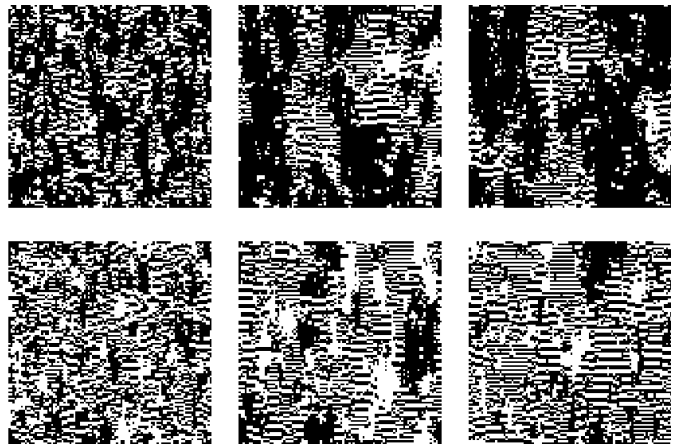}} \par}
\caption{
\label{snapshot}
Snapshots of the system starting with an uncorrelated initial state having
an initial magnetization $m_i=0.5$ (upper figures) and $m_i=0.1$ (lower figures) at times
$t=5$, $t=100$ and $t=500$, from left to the right. The four-spin interaction is given in
the vertical direction.
}
\end{figure*}

The results are shown in Fig.\ref{relax}
for different values of the initial magnetization. Here one can observe two fixed points
concerning the limiting value of $\lim_{t \to \infty} m(t) \equiv m_{\infty}$, see  Fig. \ref{relax}a
and \ref{relax}b.
For large enough initial order, $m_i>m^*$, the magnetization
approaches its finite equilibrium value in the ordered phase, $m_{\infty}=m_c$,
whereas for weak initial order, $m_i<m^*$,
the limiting magnetization is vanishing: $m_{\infty}=0$. 
According to our numerical estimates the
border between the two regimes is given by $m^*=0.500(1)$, and starting the relaxation from this
initial value the magnetization approaches a finite limiting value: $m_{\infty}=0.3844(3)$,
which is compatible with $m_c/2$. At this special value of $m_i$ there is first
a drop of the magnetization in the early times when nuclei of the coexisting phases are created (resulting in an
average magnetization of $\approx m_c/2$), which afterwards takes part in a special coarsening
process during which the coexisting ordered and disordered phases are separated and thus the interface
between the two phases are reduced. The evaluation of the structure of the system in time is shown in Fig.\ref{snapshot}. Here we present also results
for $m_i \ll m^*$, when the mass of the ordered domains and thus the magnetization in the
system is continuously decreasing.


In the regime $m_i > m^*$, the $t$ dependence of $m(t)$ is non monotonic
when $m_i$ is close to $m^*$: in the early time-steps there is a
sudden drop of the magnetization due to the formation of disordered clusters, which is then
followed by an increase of $m(t)$ that results from the formation of ordered domains. In this
regime the relaxation for $t>\tau$ is described in a power-law form:
$m(t)-m_c/2 \sim (m_i-m^*)(t-\tau)^{\theta}$. Here the measured effective exponent has a strong $m_i$
dependence and close to $m_i=m^*$ it seems to approach $\theta=0.50(3)$. Starting with the maximally ordered initial state with $m_i=1$ the relaxation of the magnetization is monotonous and
follows very closely the decay of
the autocorrelation function, $C(t)$, which is measured by preparing the system in the
fully disordered initial state, see Fig. \ref{auto}. As for the
autocorrelation function the approach of the limiting value, $m_{\infty}=m_c$, is asymptotically
given in a stretched exponential form: $m(t)-m_c \sim \exp(-b't^{a'})$, and the measured exponent is
$a'=0.196(5)$, which coincides with $a$ within the error of the calculation. We can thus conclude
that the exponents for the two functions are probably identical and they are given by: $a=a'=0.197(4)$.
Note that at a second-order transition point ($m_c=0$) the decay
with $m_i=1$ is given by $m(t) \sim t^{-x/z}$, and our result does not fit with a naive application of the
discontinuity fixed-pont value, $x=0$.
In the regime with $m_i<m^*$ and for $m_i$ close to $m^*$ the decay seems to be symmetric to the relaxation
process observed for $m_i>m^*$. On the other hand for a small $m_i/m^* \le 0.2$ the data points scales
with $m(t)/m_i$, and here we can observe also two different relaxation regions. 
As shown in Fig.\ref{relax}c, for short times, $t < \tau$,
the decay is very slow, which is well fitted by a logarithmic dependence: $m(t)/m_i \sim c-d \ln t$,
with $c=0.7$ and $d=0.1$. In the long-time limit, $t > \tau$, the decay becomes faster and
is described by a power-law form: $m(t) \sim t^{\theta'}$, with $\theta' \approx -1$, as shown in
Fig.\ref{relax}d.

\section{Discussion}
Nonequilibrium relaxation at a first-order transition point has considerable differences with
the case when the quench is performed to a second-order transition point. This is due to the
facts that i) the correlation length is finite, which leads to a finite initial time regime, $t < \tau$,
and more importantly ii) in the equilibrium state there is phase coexistence between an ordered
phase, having $m_c>0$, and a disordered phase with vanishing magnetization. In the relaxation
process the system typically evolves towards one of these phases depending on the magnetization
of the uncorrelated initial state, $m_i$. Just at $m_i=0.5$, which separates the two regimes, 
the system evolves into phase coexistence. At this special value of $m_i$ the ordered and disordered
phases are separated during a special coarsening process, thus the average magnetization approaches
$m_{\infty}=m_c/2$.
If the initial magnetization is slightly above the border-line value,
$m_i=0.5$, after the initial drop of the magnetization the ordered phase starts to grow around
the interface of the extra droplets of the ordered phase. Here we use formally the scaling
relation valid at a second-order transition point, $\theta=(x_i-x)/z$,
in which we identify the anomalous dimension of the initial magnetization with the dimension of
the interface, $x_i=d-1$, whereas we set $x=0$ at a first-order transition. Consequently we have an
initial slip exponent: $\theta=(d-1)/z$, which is compatible with the numerical results in $d=2$.
In the other case, when the initial
magnetization is small, then in the initial state there are no long-range correlations, thus the
system has a mean-field character. Since the transition point of the mean-field model is $T_c(mf)>T_c$,
an effective coarsening process takes place  during which the magnetic moments of the
domains, $\mu$, grow practically with their volume, $\mu \sim m_i \, l(t)^d \sim t^{d/z}$. As a
consequence the total magnetization of the system decays very slowly with time, slower than any power of time.
This coarsening goes on until the linear size of the
domains $l(t) \sim t^{1/z}$ approaches the correlation length $\xi$. For longer times the
magnetic moment of the domains stay constant, $\mu^* \sim m_i$, so that the magnetization
start to decay as $m(t) \sim t^{-d/z}$, which is compatible with the measured value in $d=2$.

The observed power-law dependences of the magnetization
are somewhat unexpected, since it means that
there is no relevant time scale in the problem. Thus in this sense
we have critical nonequilibrium relaxation, even though the equilibrium transition is of first order. To
understand this phenomenon we draw an analogy between nonequilibrium relaxation, i.e. the time
dependence of the magnetization, $m(t)$, after a quench from $T=\infty$ at $t=0$, and the behavior
of the equilibrium  magnetization, $m(y)$, in a semi-infinite system measured at a distance, $y$, from
the free surface located at $y=0$. In the two systems time, $t$, and distance, $y$, play analogous
roles and they can be related as $t \sim y^z$, and similarly: $\tau \sim \xi^z$. In both
systems translational invariance is broken at $t=0$ and $y=0$, respectively. At a second-order
transition point the equivalent initial-slip behaviors are given by: $m(t) \sim t^{(x_i-x)/z}$, and
$m(y) \sim y^{(x_s-x)}$, respectively, thus the anomalous dimension of the initial magnetization, $x_i$,
and the anomalous dimension of the surface magnetization, $x_s$, are analogous
quantities \cite{binder1,diehl,diehl1,ipt,pleim}. There is also an analogy when starting with a fully ordered
initial state, $m(t=0)=1$, and having an ordered surface, $m(y=0)=1$, when the asymptotic
decays are given by $m(t) \sim t^{-x/z}$ and $m(y) \sim y^{-x}$, respectively. Now let us
see how these analogies work if the transition in the system is of first order. The
properties of the surface transition in systems having a first-order transition in the bulk have
been studied in the literature \cite{lip1,lip2,kg,gk,do,ic}. Interestingly, the surface transition is of second-order for
weak surface fields, $h_s < h_s^*$, which turns to first-order for $h_s > h_s^*$, and the
two regimes are separated by a tricritical point. This phenomena for $h_s < h_s^*$ is known as surface
induced disorder and is analogous to the wetting transition \cite{diet}. Indeed in the surface region the
correlation length is divergent, and one should even define two diverging correlation lengths, $\xi_{\parallel}$
and $\xi_{\perp}$, which are related as $\xi_{\parallel} \sim \xi_{\perp}^{\zeta}$, with an anisotropy
exponent, $\zeta>1$. Now turning back to the nonequilibrium relaxation process we can interpret our
results for small initial magnetization and for $|m_i-1/2| \ll 1/2$ as due to a quench induced disordering effect, so that the nonequilibrium dynamics is critical, although the equilibrium phase transition is of first order. A closer analogy with surface induced disorder is seen if the relaxation starts with $m_i=0.5$,
where the system, after the initial short time decay, remains at the magnetization
$m_{\infty}<m_c$ for any finite times.
We note that this phenomenon is different to other kinetic analogs of
surface induced disorder\cite{kinetic} and that this type of analogy does not apply
to the case of an ordered initial state and an ordered surface. The static magnetization profile at
the ordered surface shows an exponential decay to $m_c$, whereas in the nonequilibrium relaxation problem
the decay is found to be stretched exponential.

The results obtained in this paper about the specific model in Eq.(\ref{hamilton}) are presumably
generally valid for other systems having a first-order phase transition.
In particular the functional form of the relaxation
function close to $m_i=0$ and $m_i=0.5$ are assumed to be described by universal exponents, which
could depend only on the dimension of the system. In this respect the type of dynamics (conserving and
non-conserving) does matter, e.g. for conserving dynamics the dynamical exponent should become $z=3$
instead of 2.
On the other hand the stretched exponential
form of the autocorrelation function is probably system dependent. It would be interesting
to study these questions also for other models.




\acknowledgments
This work has been supported by the Hungarian National Office of Research and
Technology under Grant No. ASEP1111, by a German-Hungarian exchange
program (DAAD-M\"OB), by the Hungarian National Research Fund under
grant No OTKA TO48721, K62588, MO45596. The simulations have been done
on Virginia Tech's System X. F.I. is indebted to L. Gr\'an\'asy for
useful discussions.


\begin{thebibliography}{0}
\bibitem{eight}
The ordered states have a strip-like structure build from $n=4$ units in the
direction of the four-spin interaction. These are $++++$, $++--$, $+-+-$, $+--+$ and
four others by interchanging $+$ and $-$. In the following we refer to the homogeneous
state with $++++$ as the ordered state.
\end{thebibliography}
\end{document}